\documentclass{osa-article}

\journal{osajournal}

\articletype{Research Article}

\usepackage{graphicx}
\usepackage{amsmath,amssymb,amsfonts}
\usepackage[caption=false]{subfig}
\usepackage[tight,nice]{units}

\usepackage{lineno}

\newcommand{\bs}{\boldsymbol}
\newcommand{\rmd}{\mathrm{d}}
\newcommand{\rmi}{\mathrm{i}}
\newcommand{\rminc}{\mathrm{inc}}
\newcommand{\rmsc}{\mathrm{sc}}
\newcommand{\rmtot}{\mathrm{tot}}
\newcommand{\rmtr}{\mathrm{tr}}
\newcommand{\rmin}{\mathrm{in}}
\newcommand{\rmout}{\mathrm{out}}
\newcommand{\rmoi}{\mathrm{oi}}
\newcommand{\rmio}{\mathrm{io}}

\begin{document}


\title{Robust Field-Only Surface Integral Equations: {S}cattering from a Dielectric Body}   

\author{Qiang Sun, \authormark{1,2,*}
  Evert Klaseboer, \authormark{3}
  Alex J.~Yuffa, \authormark{4}
  and Derek Y.~C.~Chan, \authormark{5,6}}

\address{\authormark{1} Centre of Excellence for Nanoscale BioPhotonics, RMIT University, Melbourne, VIC 3001, Australia\\
  \authormark{2} Department of Chemical Engineering, The University of Melbourne, Parkville 3010, VIC, Australia\\
  \authormark{3}  Institute of High Performance Computing, 1 Fusionopolis Way, Singapore 138632, Singapore\\
  \authormark{4} National Institute of Standards and Technology, Boulder, CO 80305 USA\\
  \authormark{5} School of Mathematics and Statistics, The University of Melbourne, Parkville 3010, VIC, Australia\\
  \authormark{6} Department of Mathematics, Swinburne University of Technology, Hawthorn VIC 3121 Australia
}

\email{\authormark{*}qiang.sun@rmit.edu.au}



\begin{abstract*}
A robust and efficient field-only nonsingular surface integral method to solve Maxwell's equations for the components of the electric field on the surface of a dielectric scatterer is introduced. In this method, both the vector Helmholtz equation and the divergence-free constraint are satisfied inside and outside the scatterer. The divergence-free condition is replaced by an equivalent boundary condition that relates the normal derivatives of the electric field across the surface of the scatterer. Also, the continuity and jump conditions on the electric and magnetic fields are expressed in terms of the electric field across the surface of the scatterer. Together with these boundary conditions, the scalar Helmholtz equation for the components of the electric field inside and outside the scatterer is solved by a fully desingularized surface integral method. Comparing with the most popular surface integral methods based on the Stratton--Chu formulation or the PMCHWT formulation, our method is conceptually simpler and numerically straightforward because there is no need to introduce intermediate quantities such as surface currents and the use of complicated vector basis
functions can be avoided altogether. Also, our method is not affected by numerical issues such as the zero frequency catastrophe and does not contain integrals with (strong) singularities.  To illustrate the robustness and versatility of our method, we show examples in the Rayleigh, Mie, and geometrical optics scattering regimes. Given the symmetry between the electric field and the magnetic field, our theoretical framework can also be used to solve for the magnetic field.
\end{abstract*}


\section{\label{sec:intro} Introduction}
There have been two recent independent developments in formulating computational electromagnetics (CEM) scattering~\cite{FrezzaJOSA2018} in terms of surface integral equations~\cite{Yuffa2018,Klaseboer2017} that are conceptually very different from the venerable theoretical framework of Stratton--Chu which was established almost 80 years ago~\cite{Stratton1939, Stratton1941} or the PMCHWT formulation \cite{Poggio1973, Wu1977, Chang1977} or the potential based CEM methods~\cite{Vico2016, Fu2017}. These earlier methods either entail solving for surface currents or charges at boundaries or for the scalar and vector potentials, whereas the recent works are based on solving directly for components of the electric field. One of the field-only formulations had its genesis in the study of scattering from (i) infinite rough surfaces~\cite{DeSanto1993} some 25 years ago, (ii) finite dielectric bodies~\cite{Yuffa2006} more than a decade ago, and has been recently generalized with an extensive use of differential geometry~\cite{Yuffa2018}.  The other field-only formulation~\cite{Klaseboer2017, Sun2017} focused on the use of nonsingular surface integral equations for the field components.  This method stems from an observation that the physical phenomena is finite and well-behaved on boundaries, and thus should not contain mathematically singular kernels.  In this method, the  divergence-free condition was satisfied via the identity 
\begin{equation*}
  \nabla^2 (\bs{r} \cdot \bs{E}) + k^2 (\bs{r} \cdot \bs{E})= 2\bs{\nabla} \cdot \bs{E} = 0,
\end{equation*}
where $\bs r$ the position vector, and resulted in an additional Helmholtz equation for $\bs{r} \cdot \bs{E}$ that led to a $9N \times 9N$ system of linear equations~\cite{Sun2017}.

In this paper, we combine the above two field-only integral methods to obtain a nonsingular integral formulation, which when discretized yields a $6N \times 6N$ system of linear equations.  Therefore, the framework developed in this paper gives a $56\%$ reduction in memory requirements and subsequently leads to faster solution times.  Furthermore, this approach turns out to be conceptually simple and can  provide direct access to values of the field and its normal derivatives on the boundary of the scatterer. The implementation is free of mathematical singularities and facilitates the use of simple, efficient and accurate surface integration algorithms.  It also should be noted that this paper is a natural generalization of our previous publication~\cite{SunFuturePartI}.  In~\cite{SunFuturePartI}, we considered the much simpler case of scattering by a perfect electric conductor (PEC) in order not to obscure the conceptual simplicity and elegance of the method by the non-zero internal fields.

The paper is organized as follows.  The theoretical framework of our formulation is explained in Section~\ref{sec:formulation}.  In Section~\ref{sec:result}, we consider numerical examples of interest to the optics community in the Rayleigh, Mie, and geometric optics scattering regimes.   Finally, some concluding remarks are presented in Section~\ref{sec:CONCLUDE} as well as a prescription how to modify our formulation if the magnetic fields are of primary interest.

\section{\label{sec:formulation} Field-Only Formulation}
In a source-free, linear, homogeneous medium the propagation of a time-harmonic electric field $\bs{E}(\bs{r}) \exp(-\rmi \omega t)$, with $t$ denoting time and $\omega$ denoting the angular frequency, is governed by the vector Helmholtz equation
\begin{equation} \label{eq:EH_wave_eqn}
\nabla^2 \bs{E}(\bs{r}) + k^2 \bs{E}(\bs{r}) = \bs{0},
\end{equation}
where $k =  \sqrt{\epsilon \mu} \omega$ is the wavenumber with $\epsilon$ and $\mu$ being the permittivity and permeability of the medium, respectively. Thus, each Cartesian component of $\bs{E}$ satisfies the scalar Helmholtz wave equation
\begin{equation} \label{eq:E_scalar_wave_eqn}
\nabla^2 E_\alpha + k^2 E_\alpha = 0, \qquad \alpha = x, y, z.
\end{equation}
The electric field is also divergence-free, i.e.,
\begin{equation} \label{eq:div_E_zero}
\bs{\nabla} \cdot \bs{E} = 0,   
\end{equation}
thus, in principle there are only two independent components of $\bs{E}$ that have to be determined.

In a typical scattering problem, an incident wave, $\bs{E}^{\rminc}$, is scattered by a dielectric body, and the resulting scattered field outside the scatterer as well as the transmitted field inside the scatterer are to be determined.  After accounting for the fact that the scattered field, $\bs{E}^{\rmsc}$, obeys the Silver--M\"{u}ller radiation condition~\cite{Colton1992}, the transmitted field, $\bs{E}^{\rmtr}$, is finite inside the scatterer, and both $\bs{E}^\rmsc$ and $\bs{E}^\rmtr$ satisfy \eqref{eq:div_E_zero}, we see that there are only four unknown \emph{scalar} functions (two for each domain).  These functions are usually found by solving \eqref{eq:E_scalar_wave_eqn} and applying the continuity conditions for the tangential components of the electric and magnetic fields.  In our formulation, the key point of departure from the formulation outlined above is to cast the divergence-free condition in the 3D domain as a boundary condition.  Since the problem is elliptic in nature, this should always be possible. Casting the divergence-free condition as a boundary condition enables us to directly solve for the components of $\bs{E}$.  Furthermore, it guarantees that $\bs{E}^\rmsc$ ($\bs{E}^\rmtr$) satisfies the divergence-free condition in the 3D domain outside (inside) the scatterer \cite{Yuffa2018,SunFuturePartI}.

The value of $\bs{\nabla} \cdot \bs{E}$ on the scatterer's surface $S$ can be expressed using differential geometry as a combination of the normal component of $\bs{E}$ on $S$ and the normal as well as the tangential derivatives of $\bs{E}$ on $S$ (see \cite[equation (A12)]{SunFuturePartI} or \cite[equation (23)]{Yuffa2018}). That is, at any point on the surface $S$, we have
\begin{equation}\label{eq:divE_zero_surf}
  \bs{\nabla} \cdot \bs{E} =  \bs{n} \cdot \frac{\partial \bs{E} } { \partial{n} }
  - \kappa E_{n} + \frac{\partial { E_{t_{1}}  }} { \partial{t_{1}} }
  + \frac{\partial { E_{t_{2}}  }} { \partial{t_{2}} }=0,
\end{equation}
where $\kappa$ is the mean curvature.  In \eqref{eq:divE_zero_surf}, $\bs{n}$ is the unit normal pointing into the scatterer, $E_n = \bs{n} \cdot \bs{E}$ is the normal component of $\bs{E}$, and $E_{t_{1}} = \bs{t}_1 \cdot \bs{E}$ and $E_{t_{2}} = \bs{t}_2 \cdot \bs{E}$ are the tangential components of $\bs{E}$ along the two mutually perpendicular tangential unit vectors $\bs{t}_1$ and $\bs{t}_2$. The normal and tangential derivatives are defined by $\partial(\cdot) / \partial n = \bs{n} \cdot \bs{\nabla}(\cdot)$ and $\partial(\cdot) / \partial t_j = \bs{t}_{j} \cdot \bs{\nabla}(\cdot)$ for $j=1,2$, respectively.

We will use \eqref{eq:divE_zero_surf} to decompose the standard surface integral representation written for the three Cartesian components of the electric field into its normal and tangential components. Outside the scatterer we use Green's second identity to express the solution of \eqref{eq:E_scalar_wave_eqn} for the scattered field, $\bs{E}^{\rmsc}(\bs{r}_0)$, in terms of integrals over the surface values 
\begin{subequations}
  \label{eq:BIE_E_components} 
  \begin{multline}\label{eq:BIE_E_components_a} 
    c_0(\bs{r}_0)  E^{\rmsc}_\alpha(\bs{r}_0)
    + \int_S E^{\rmsc}_\alpha(\bs{r})
    \frac{\partial G(\bs{r}, \bs{r}_0)}{\partial n} \, \rmd S(\bs{r}) 
    = \int_S  \frac{\partial E^{\rmsc}_\alpha(\bs{r})}{\partial n}
    G(\bs{r}, \bs{r}_0) \, \rmd S(\bs{r}), \quad \alpha = x, y, z,
  \end{multline}
where $c_0 = 4 \pi$ if $\bs{r}_0 \notin S$ (i.e., when $\bs{r}_0$ is in the 3D domain outside the scatterer).  If $\bs{r}_0 \in S$ (approached from the exterior 3D domain), then $c_0$ is the solid angle subtended at $\bs{r}_0$.  The integral representation of the transmitted field, $\bs{E}^\rmtr$, inside the scatterer is given by
\begin{multline}\label{eq:BIE_E_components_b} 
    c_0^{\rmin}(\bs{r}_0)  E^{\rmtr}_\alpha(\bs{r}_0)
    - \int_S E^{\rmtr}_\alpha(\bs{r})
    \frac{\partial G_{\rmin}(\bs{r}, \bs{r}_0)}{\partial n} \, \rmd S(\bs{r}) 
    = -\int_S  \frac{\partial E^{\rmtr}_\alpha(\bs{r})}{\partial n}
    G_{\rmin}(\bs{r}, \bs{r}_0) \, \rmd S(\bs{r}), \quad \alpha = x, y, z, 
  \end{multline}
\end{subequations}
where $\bs{r}_0$ is inside the scatterer.  Equation \eqref{eq:BIE_E_components_b} follows directly from the application of Green's second identity to \eqref{eq:E_scalar_wave_eqn} with $E_\alpha = E_{\alpha}^\rmtr$ and the two minus signs appear because the normal vector points into the scatterer. In \eqref{eq:BIE_E_components_b}, $c_0^{\rmin} = 4 \pi$ if $\bs{r}_0 \notin S$ (i.e., $\bs{r}_0$ is inside the scatterer) and $c_0^{\rmin}$ is the solid angle if $\bs{r}_0 \in S$ when $\bs{r}_{0}$ approaches to $S$ from inside the scatterer. It is worth mentioning that $c_0 + c_0^{\rmin} = 4\pi$ if $\bs{r}_0 \in S$. In \eqref{eq:BIE_E_components}, Green's function is $G(\bs{r},\bs{r}_0)= \exp(\rmi k |\bs{r} - \bs{r}_0| / |\bs{r} - \bs{r}_0|,$ where $k$ denotes the appropriate wavenumber for the region, i.e., $k= k_\rmin = \sqrt{\epsilon_\rmin \mu_\rmin} \omega $  for $G_{\rmin}(\bs{r}, \bs{r}_0)$ (inside the scatterer) or $ k=k_\rmout = \sqrt{\epsilon_\rmout \mu_\rmout} \omega$ for $G(\bs{r}, \bs{r}_0)$ (outside the scatterer) .

At this point in the formulation, we see that \eqref{eq:BIE_E_components} contains $12$ unknown functions on $S$; namely, $\{E_{\alpha}^{\rmsc}, \partial E_{\alpha}^{\rmsc}/\partial n \}$ and $\{E_{\alpha}^{\rmsc}$, $\partial E_\alpha^{\rmsc}/\partial n\}$, $\alpha = x, y, z$.  In order to determine the $12$ unknown functions we need 12 equations.  Six of these equations come from \eqref{eq:BIE_E_components}. Three more equations come from the continuity conditions satisfied by the electric field on $S$, namely,
\begin{subequations}\label{eq:En_BC_and_Et_BC}
\begin{equation} \label{eq:En_BC}
  E_{n}^{\rmtr} = \epsilon_{\rmoi} (E_{n}^{\rminc}+E_{n}^{\rmsc}), \quad \epsilon_{\rmoi} \equiv
  \epsilon_{\rmout} / \epsilon_{\rmin}
\end{equation}
and
\begin{equation} \label{eq:Et_BC}
  E_{t_j}^\rminc + E_{t_j}^\rmsc = E_{t_j}^\rmtr, \quad  j=1,2 . 
\end{equation}
\end{subequations}
The last three equations come from the continuity condition satisfied by the normal derivative of the electric field, $\partial \bs{E}/\partial n$, on $S$.

To derive these last three equations, we write \eqref{eq:divE_zero_surf} for the total exterior field, $\bs{E}^\rmsc + \bs{E}^\rminc$,  and subtract the corresponding equation for the transmitted field, $\bs{E}^\rmtr$. Then, after using \eqref{eq:En_BC_and_Et_BC}, we obtain
\begin{subequations}\label{eq:dEdn_full}
\begin{equation} \label{eq:dEdn_n_BC}
  \bs{n} \cdot \frac{\partial { \bs{E}^{\rmtr} }  }{ \partial{n} } =  \kappa (\epsilon_{\rmoi}-1) \left(E^{\rmsc}_{n}+E^{\rminc}_{n} \right) +  \bs{n} \cdot \frac{\partial { \bs{E}^{\rmsc} }  }{ \partial{n} } +   \bs{n} \cdot \frac{\partial { \bs{E}^{\rminc} }  }{ \partial{n} } .
\end{equation}
Equation \eqref{eq:dEdn_n_BC} only provides a continuity condition for the \emph{normal} component of $\partial \bs{E}/\partial n$.  To obtain a continuity condition for the tangential components of $\partial \bs{E}/\partial n$, we express the continuity condition for the tangential components of $\bs{H}$ on $S$, i.e.,
\begin{equation}\label{eq:Ht_BC}
H_{t_j}^\rminc + H_{t_j}^\rmsc = H_{t_j}^\rmtr \quad \text{for}\quad j=1,2,
\end{equation}
in terms of the electric field to obtain (see \ref{Appendix_HnHt} for details)
\begin{multline}\label{eq:dEdn_t_BC}
  \left(\epsilon_{\rmoi}- \mu_{\rmio} \right) \frac{\partial}{\partial t_j} \left[E_n^{\rminc} + E_n^{\rmsc}\right]
  + \kappa_j\left(1-\mu_{\rmio}\right)  \left[E_{t_j}^{\rminc} + E_{t_j}^{\rmsc}\right] \\
  + \mu_{\rmio} \left(\bs{t}_j \cdot \frac{\partial \bs{E}^{\rminc}}{\partial n} 
  +\bs{t}_j \cdot \frac{\partial \bs{E}^{\rmsc} }{\partial n} \right)
  =  \bs{t}_j \cdot \frac{\partial \bs{E}^{\rmtr}}{\partial n}, 
\end{multline}
\end{subequations}
where $\mu_{\rmio}\equiv \mu_{\rmin}/\mu_{\rmout}$, $\kappa_j$ is the local curvature along the $\bs{t}_j$ direction and $j=1,2$.   In the limit $\mu_{\rmio}=1$, \eqref{eq:dEdn_t_BC} reduces to
\begin{equation}\label{eq:tE_nonmagnetic}
  \left(1-\epsilon_{\rmoi}^{-1}\right) \frac{\partial E_n^\rmtr}{\partial t_j}
  = \bs{t}_j \cdot \frac{\partial}{\partial n}
  \left[\bs{E}^\rmtr - \left(\bs{E}^\rminc + \bs{E}^\rmsc\right)  \right]
\end{equation}
for $j=1,2$.  Equation \eqref{eq:tE_nonmagnetic} states that in a nonmagnetic medium the \emph{tangential components} of the \emph{normal derivative} of the electric field are discontinuous across an interface by an amount proportional to the \emph{tangential derivative} of the \emph{normal component} of the electric field inside the scatterer.  Furthermore, if there is no scatterer, i.e., $\epsilon_\rmoi = 1$, then \eqref{eq:tE_nonmagnetic} reduces to the expected form; namely,  
$\bs{t}_j \cdot \frac{\partial}{\partial n} \left(\bs{E}^\rminc + \bs{E}^\rmsc\right) = \bs{t}_j \cdot \frac{\partial}{\partial n}\bs{E}^\rmtr$ for $j=1,2$.

Lastly, we note that \eqref{eq:En_BC_and_Et_BC} and \eqref{eq:dEdn_full} are simply equations (9) and (19) in \cite{Yuffa2018}, respectively, written in a different notation.  Furthermore, \eqref{eq:dEdn_full} (or equivalently equation (19) in \cite{Yuffa2018}) is not widely known to the scientific community but is an essential equation for our surface integral method.

\subsection{Numerical Solution}\label{sec:theory_implementation}
One approach to obtain a numerical solution is to directly discretize the surface integral equations given by \eqref{eq:BIE_E_components} \cite{Yuffa2019}.  This approach will yield a system of linear equations that can be solved for the chosen unknowns: $\{E_{n}^{\rmsc},  E_{t_{1}}^{\rmsc}, E_{t_{2}}^{\rmsc}\}$ and  $\{\bs{n} \cdot \partial { \bs{E}^{\rmsc} }  / {\partial{n}}, \bs{t}_{1} \cdot \partial { \bs{E}^{\rmsc} }  / {\partial{n}} , \bs{t}_{2} \cdot \partial { \bs{E}^{\rmsc} }  / {\partial{n}} \}$.  Unfortunately, this approach requires the discretization of singular kernels (Green's function and its normal derivative) and, therefore, much care must be taken to avoid numerical difficulties~\cite{Yuffa2019}.  Another approach would be to use our recently developed robust and accurate desingularized method~\cite{Klaseboer2017, Sun2017}, where the singular behavior of Green's function and its normal derivative is ``subtracted out'' before the discretization.  This is the method we have chosen to use here and it is explained in more detail in \ref{Appendix_desingularized} (also see \cite{SunFuturePartI}).

From \ref{Appendix_desingularized}, we see that the nonsingular version of \eqref{eq:BIE_E_components} is given by
\begin{multline} \label{eq:NS_BIE_sigma}
 \int_{\Sigma} { \left[\frac{\partial {p (\bs{r})}} {{\partial {n}}} - p(\bs{r}_0) \frac{\partial {g (\bs{r})}} {{\partial {n}}} - \frac{\partial {p(\bs{r}_0)}} {{\partial {n}}} \frac{\partial {f (\bs{r})}} {{\partial {n}}} \right] G \, \rmd S(\bs{r}}) \\
=\int_{\Sigma} {\left[p(\bs{r}) - p(\bs{r}_0) g(\bs{r}) - \frac{\partial {p(\bs{r}_0)}} {{\partial {n}}}  f(\bs{r})\right] \frac{\partial {G}} {{\partial {n}}} \, \rmd S(\bs{r}}),  
\end{multline}
where $f$ and $g$ are auxiliary functions that ``subtract out'' the singular behavior of the kernels.  For the interior problem, $p$ is one of the Cartesian components of the transmitted field, i.e., $p = E_{\alpha}^{\rmtr}, \alpha = x,y,z$, and $\Sigma = S$.  Similarly, for the exterior problem, $p$ is one of the Cartesian components of the scattered field but $\Sigma = S + S_{\infty}$, where $S_{\infty}$ is an artificial sphere of infinite radius.  Note that the contribution from $S_{\infty}$ is generally non-zero because $f$ and $g$ may not decay as fast as the scattered field at infinity.  However, with our choice of $f$ and $g$ the integrals over $S_{\infty}$ may be performed analytically, and thus are not of much concern, see \ref{Appendix_desingularized}.

For the exterior problem, after discretizing the surface $S$ into six-noded quadratic triangular elements~\cite{Klaseboer2017, Sun2017}, the surface integral equation \eqref{eq:NS_BIE_sigma} is converted into a surface element matrix system connecting all $N$ nodes to their normal derivatives via
\begin{subequations}\label{eq:BIE_GH_Matrix}
\begin{equation} \label{eq:BIE_GH_Matrix_in} 
 {\cal{H}} \cdot \underline{p^{\rmsc}} =  {\cal{G}} \cdot \underline{\frac{\partial p^{\rmsc}}{\partial n}}
\end{equation}
In \eqref{eq:BIE_GH_Matrix_in}, $\underline{p^{\rmsc}} = \underline{E^{\rmsc}_{\alpha}}$ (with $\alpha=x, y \text{ or } z$) represents a column vector with all of the $N$ node values of $p^{\rmsc}$, $\underline{\frac{\partial p^{\rmsc}}{\partial n}}$ is a similar column vector for the normal derivatives of $p^{\rmsc}$. For explicit examples of $\cal{G}$ and $\cal{H}$ see Appendix~B in \cite{SunFuturePartI}. Another matrix system can be constructed for the transmitted field (interior problem) but with ${\cal{H}}_{\rmin}$ and ${\cal{G}}_{\rmin}$ matrices which can also be obtained following the same procedure demonstrated in Appendix~B in \cite{SunFuturePartI}.  These matrices differ from ${\cal{H}}$ and ${\cal{G}}$ because ${\cal{H}}_{\rmin}$ and ${\cal{G}}_{\rmin}$ do not contain contributions from integrals over $S_{\infty}$ and the Green's function inside the scatterer has a different wavenumber $k$.  For completeness and to facilitate the development that follows, we explicitly write this relationship as
\begin{equation} \label{eq:BIE_GH_Matrix_out} 
 {\cal{H}}_{\rmin} \cdot \underline{p^{\rmtr}} =  {\cal{G}}_{\rmin} \cdot \underline{\frac{\partial p^{\rmtr}}{\partial n}},
\end{equation}
\end{subequations}
where $\underline{p^{\rmtr}} = \underline{E^{\rmtr}_{\alpha}}$ and $\alpha=x, y \text{ or } z$.

We need to use the boundary conditions given by \eqref{eq:En_BC_and_Et_BC} and \eqref{eq:dEdn_full} to eliminate $\underline{p^{\rmtr}}$ and $\underline{\frac{\partial p^{\rmtr}}{\partial n}}$ from \eqref{eq:NS_BIE_sigma}.  However, the boundary conditions are written in terms of the normal and tangential components and \eqref{eq:NS_BIE_sigma} requires the Cartesian components.  To reconcile this mismatch, we project the normal and tangential basis onto the Cartesian basis $\{\bs{e}_x, \bs{e}_y, \bs{e}_z\}$, i.e.,
\begin{subequations}\label{eq:E_dEdn_in_xyz}
  \begin{equation}\label{eq:E_in_xyz}
    E_{\alpha} \bs{e}_{\alpha} = \left[ n_{\alpha} E_n \right] \bs{e}_{\alpha} + \left[ t_{1\alpha} E_{t_1} \right] \bs{e}_{\alpha}
    + \left[ t_{2\alpha} E_{t_2} \right] \bs{e}_{\alpha},
  \end{equation}
  \begin{multline}\label{eq:dEdn_in_xyz}
    \frac{ \partial E_{\alpha}}{\partial n} \bs{e}_{\alpha}
    = \left[n_{\alpha} \left(\bs{n} \cdot
        \frac{\partial \bs{E}}{\partial n} \right)\right] \bs{e}_{\alpha}
    + \left[t_{1\alpha} \left(\bs{t}_1 \cdot
        \frac{\partial \bs{E}}{\partial n} \right)\right] \bs{e}_{\alpha} 
    + \left[t_{2\alpha} \left(\bs{t}_2 \cdot
        \frac{\partial \bs{E}}{\partial n} \right)\right] \bs{e}_{\alpha},
    \quad \alpha = x, y, z,
  \end{multline}
\end{subequations}    
where $n_{\alpha} = \bs{n} \cdot \bs{e}_{\alpha}$, $t_{1\alpha} = \bs{t}_1 \cdot \bs{e}_{\alpha}$, $t_{2\alpha} = \bs{t}_2 \cdot \bs{e}_{\alpha}$, and $E$ denotes $E^{\rmsc}$ or $E^{\rmtr}$.  Finally, using \eqref{eq:E_dEdn_in_xyz} and the boundary conditions at all of the nodes on the surface, we obtain $6N \times 6N$ system of linear equations for the chosen boundary unknowns.  This linear system is given by
%
\begin{subequations}\label{eq:big_linear_system_full}
  \begin{equation}\label{eq:big_linear_system}
    \begin{bmatrix}
      n_{x}{\cal{H}} & t_{1x}{\cal{H}} & t_{2x}{\cal{H}}
      & - {n_{x}}{\cal{G}} & - {t_{1x}}{\cal{G}}
      & - {t_{2x}}{\cal{G}} \\
      n_{y}{\cal{H}} & t_{1y}{\cal{H}} & t_{2y}{\cal{H}}
      & - {n_{y}}{\cal{G}} & - {t_{1y}}{\cal{G}}
      & - {t_{2y}}{\cal{G}} \\
      n_{z}{\cal{H}} & t_{1z}{\cal{H}} & t_{2z}{\cal{H}}
      & - {n_{z}}{\cal{G}} & - {t_{1z}}{\cal{G}}
      & - {t_{2z}}{\cal{G}} \\
       \bar{{\cal{H}}}^{n_{x}}_{\rmin}  & \bar{{\cal{H}}}^{t_{1x}}_{\rmin}
      &   \bar{{\cal{H}}}^{t_{2x}}_{\rmin}  & -n_{x}{\cal{G}}_{\rmin}
      & - \mu_{\rmio}t_{1x} {\cal{G}}_{\rmin} & -\mu_{\rmio} t_{2x} {\cal{G}}_{\rmin}  \\
      \bar{{\cal{H}}}^{n_{y}}_{\rmin}  & \bar{{\cal{H}}}^{t_{1y}}_{\rmin}
      &   \bar{{\cal{H}}}^{t_{2y}}_{\rmin}  &  -n_{y}{\cal{G}}_{\rmin}
      & - \mu_{\rmio}t_{1y} {\cal{G}}_{\rmin} & -\mu_{\rmio} t_{2y} {\cal{G}}_{\rmin}  \\ 
      \bar{{\cal{H}}}^{n_{z}}_{\rmin}  & \bar{{\cal{H}}}^{t_{1z}}_{\rmin}
      &   \bar{{\cal{H}}}^{t_{2z}}_{\rmin}  &  -n_{z}{\cal{G}}_{\rmin}
      & - \mu_{\rmio}t_{1z} {\cal{G}}_{\rmin} & -\mu_{\rmio} t_{2z} {\cal{G}}_{\rmin}
    \end{bmatrix}
    \begin{bmatrix}
      E_n^{\rmsc} \\ E_{t_{1}}^{\rmsc} \\ E_{t_{2}}^{\rmsc} \\
      \bs{n} \cdot \frac{\partial \bs{E}^{\rmsc}}{\partial n}   \\
      \bs{t}_{1} \cdot \frac{\partial \bs{E}^{\rmsc}}{\partial n}  \\
      \bs{t}_{2} \cdot \frac{\partial \bs{E}^{\rmsc}}{\partial n}
    \end{bmatrix}
    =
    \begin{bmatrix}
      0 \\ 0 \\ 0 \\
      {\cal{B}}_{x} \\ {\cal{B}}_{y} \\ {\cal{B}}_{z}
    \end{bmatrix}, 
  \end{equation}
  where
  \begin{align}
    \bar{{\cal{H}}}^{n\alpha}_{\rmin} &= \epsilon_{\rmoi} n_{\alpha} {\cal{H}}_{\rmin}
                                - \kappa \left(\epsilon_{\rmoi}-1\right)
                                n_{\alpha} {\cal{G}}_{\rmin}
                                - (\epsilon_{\rmoi}-\mu_{\rmio})
                                t_{1\alpha}{\cal{G}}_{\rmin}
                                \frac{\partial }{\partial t_1}
                                - (\epsilon_{\rmoi}-\mu_{\rmio})
                                t_{2\alpha}  {\cal{G}}_{\rmin}
                                \frac{\partial}{\partial t_2}, \label{eq:matrix_patial_derivative} \\
    \bar{{\cal{H}}}^{t1\alpha}_{\rmin} &= t_{1\alpha} {\cal{H}}_{\rmin}
                                 -  \kappa_{1}(1-\mu_{\rmio})
                                 t_{1\alpha}  {\cal{G}}_{\rmin} \quad \text{and} \quad
                                 \bar{{\cal{H}}}^{t2\alpha} =   t_{2\alpha}
                                 {\cal{H}}_{\rmin} - \kappa_{2} (1-\mu_{\rmio})
                                 t_{2\alpha}  {\cal{G}}_{\rmin},
  \end{align}
  and
  \begin{align}
    {\cal{B}}_{\alpha} =& - \epsilon_{\rmoi} n_{\alpha} {\cal{H}}_{\rmin}     E_n^{\rminc} - t_{1\alpha} {\cal{H}}_{\rmin}E^{\rminc}_{t_{1}} - t_{2\alpha}{\cal{H}}_{\rmin}E^{\rminc}_{t_{2}} \nonumber \\
                        & + n_{\alpha}{\cal{G}}_{\rmin} \left[\kappa (\epsilon_{\rmoi}-1) E^{\rminc}_{n} + \bs{n} \cdot \frac{\partial { \bs{E}^{\rminc} }  }{ \partial{n} }\right] \nonumber \\
                       &+ t_{1\alpha}{\cal{G}}_{\rmin}
                         \left[(\epsilon_{\rmoi}-\mu_{\rmio})
                         \frac{\partial E^{\rminc}_{n}}{\partial t_1}
                         + \kappa_1(1-\mu_{\rmio})E^{\rminc}_{t_1}
                         + \mu_{\rmio} \bs{t}_1 \cdot
                         \frac{\partial \bs{E}^{\rminc}}{\partial n} \right]
                         \nonumber \\
                        & + t_{2\alpha}{\cal{G}}_{\rmin}
                          \left[(\epsilon_{\rmoi}-\mu_{\rmio})
                          \frac{\partial E^{\rminc}_{n}}{\partial t_2}
                          + \kappa_2(1-\mu_{\rmio})E^{\rminc}_{t_{2}}
                          + \mu_{\rmio} \bs{t}_{2} \cdot
                          \frac{\partial \bs{E}^{\rminc}}{\partial n}\right]
  \end{align}
  with $\alpha = x,y,z.$
\end{subequations}

The assembly of \eqref{eq:big_linear_system_full} is straightforward, except perhaps for the last three terms in the first column of \eqref{eq:big_linear_system} because they contain tangential partial derivatives, see \eqref{eq:matrix_patial_derivative}. We explain the numerical implementation of these tangential derivatives as well as the derivatives that are used to calculate the curvatures $\kappa_1$ and $\kappa_2$ in \ref{Appendix_dtMatrix}. When the $6N \times 6N$ matrix system of \eqref{eq:big_linear_system} is compared to the $9N \times 9N$ matrix system in \cite{Sun2017}, it is clear that the memory required is reduced by 56\% ($9^2$ vs. $6^2$). Furthermore, the $9N \times 9N$ matrix system contained many zero entries, whereas \eqref{eq:big_linear_system} is a full matrix system.  

If there is no scatterer, i.e., a transparent object, then $k_{\rmin}=k_{\rmout}$, $\mu_{\rmio}=1$, $\epsilon_{\rmoi}=1$, ${\cal{G}}_{\rmin} ={\cal{G}}$ but ${\cal{H}}_{\rmin}$ and ${\cal{H}}$ differ by a factor $4\pi$ on the diagonal.  In this case, we see that \eqref{eq:big_linear_system_full} yields the expected solution; namely, $\bs{E}^{\rmsc}=0$ and consequently $\bs{E}^{\rmtr} = \bs{E}^{\rminc}$. In \cite{Sun2017}, it was also shown that this framework applied to planar dielectrics reverts back to the Fresnel equations and Snell's law. 

If the scatterer is a perfect electric conductor (PEC), $\mu_{\rmio}=1$ and $\epsilon_{\rmoi} \to 0$ as the imaginary part of $\epsilon_{\rmin}$ goes to infinity, then only the fields outside of the PEC scatterer are nonzero and on the boundary of the PEC scatterer the tangential components of the total electric field, $\bs{E}^{\rminc} + \bs{E}^{\rmsc}$, vanish. Furthermore,  \eqref{eq:dEdn_n_BC} reduces to
\begin{equation} \label{eq:PEC_div0}
\bs n \cdot \frac{\partial \bs{E}^\rmsc}{\partial n} - \kappa E_n^\rmsc = -\bs{n} \cdot \frac{\partial \bs{E}^\rminc}{\partial n} + \kappa E_n^\rminc,
\end{equation}
which agrees with our previous result, see equation (12) in \cite{SunFuturePartI}.  Also, it can be shown that the first three rows of \eqref{eq:big_linear_system} reduce to a $3 N \times 3 N$ linear system which is the same as  equation (13) of our previous paper~\cite{SunFuturePartI} where a more detailed discussion of the PEC case can be found. It is also instructive to exhibit the limiting forms of the last three rows of \eqref{eq:big_linear_system} when $\epsilon_{\rmoi}=0$ and $\mu_{\rmio}=1$. For example, in this limit, the fourth row reduces to%
\begin{equation} \label{eq:Matrix_last3rowsPEC}
    \begin{aligned}
    {\cal{G}}_{\rmin}\Bigg\{ & n_x \left[\kappa E_n^{\rmsc}  - \bs n \cdot \frac{\partial \bs{E}^\rmsc}{\partial n} +\kappa E_n^{\rminc} - \bs n \cdot \frac{\partial \bs{E}^\rminc}{\partial n}\right]
    \\
     &+t_{1x} \left[ \frac{\partial \bs{E}_n^\rmsc}{\partial t_1} - \bs t_1 \cdot \frac{\partial \bs{E}^\rmsc}{\partial n} + \frac{\partial \bs{E}_n^\rminc}{\partial t_1} - \bs t_1 \cdot \frac{\partial \bs{E}^\rminc}{\partial n}\right]
    \\
     &+t_{2x} \left[ \frac{\partial \bs{E}_n^\rmsc}{\partial t_2}  - \bs t_2 \cdot \frac{\partial \bs{E}^\rmsc}{\partial n} + \frac{\partial \bs{E}_n^\rminc}{\partial t_2} - \bs t_2 \cdot \frac{\partial \bs{E}^\rminc}{\partial n}\right]\Bigg\}
     \\
      =-{\cal{H}}_{\rmin}&\left[t_{1x} E_{t1}^{\rmsc} + t_{2x} E_{t2}^{\rmsc}+ t_{1x} E_{t1}^{\rminc}+ t_{2x} E_{t2}^{\rminc} \right].
    \end{aligned}
\end{equation}
The first line in the above equation is just \eqref{eq:dEdn_n_BC} with $\bs{E}^{\rmtr}=\bs{0}$, $\epsilon_{\rmoi}=0$, and $\mu_{\rmio}=1$. The second and third lines are \eqref{eq:dEdn_t_BC} for $\bs{t}_1$ and $\bs{t}_2$, respectively. The last line becomes zero because the tangential components of the total electric field vanish on the PEC surface. All terms in brackets are now zero. A similar derivation can be carried out for the fourth and fifth row of \eqref{eq:big_linear_system}. 
Thus, the constructed matrix system is self-consistent and reverts back to the correct physical limits for a transparent or a PEC object. 

Finally, \eqref{eq:big_linear_system} can be solved numerically to obtain values of $\bs{E}^{\rmsc}$ and $\partial \bs{E}^{\rmsc}/\partial n $ on the surface of the scatterer and subsequently, values of $\partial \bs{E}^{\rmtr}/\partial n $ and $\bs{E}^{\rmtr}$ on $S$ can be found by post-processing. $\bs{E}^{\rmtr}$ can be found by using \eqref{eq:Et_BC} and \eqref{eq:En_BC}, and $\partial \bs{E}^{\rmtr}/\partial n $ can be obtained by using (\ref{eq:dEdn_n_BC}) and (\ref{eq:dEdn_t_BC}).  Thus, after the post-processing, we have all electric fields and their normal derivatives on $S$ and therefore, we can compute the electric field anywhere inside and outside the scatterer.  For example, this can be done via the formulation given in~\cite{Sun2015} so that the numerical results are not affected by the near singular nature of \eqref{eq:BIE_E_components}. Although we have chosen to work with the exterior field's boundary unknowns, i.e., $\bs{E}^\rmsc$ and $\partial \bs{E}^\rmsc/\partial n$, it is also equally valid to choose the interior field's boundary unknowns, i.e., $\bs{E}^{\rmtr}$ and $\partial \bs{E}^{\rmtr}/\partial n $.  This choice may be of interest in photonics applications and is further discussed in \ref{Appendix_MatrixEtr}.

\section{\label{sec:result} Results}%
\begin{figure*}[!t]
\centering{}
\subfloat[(a)]{ \includegraphics[width=0.6\textwidth]{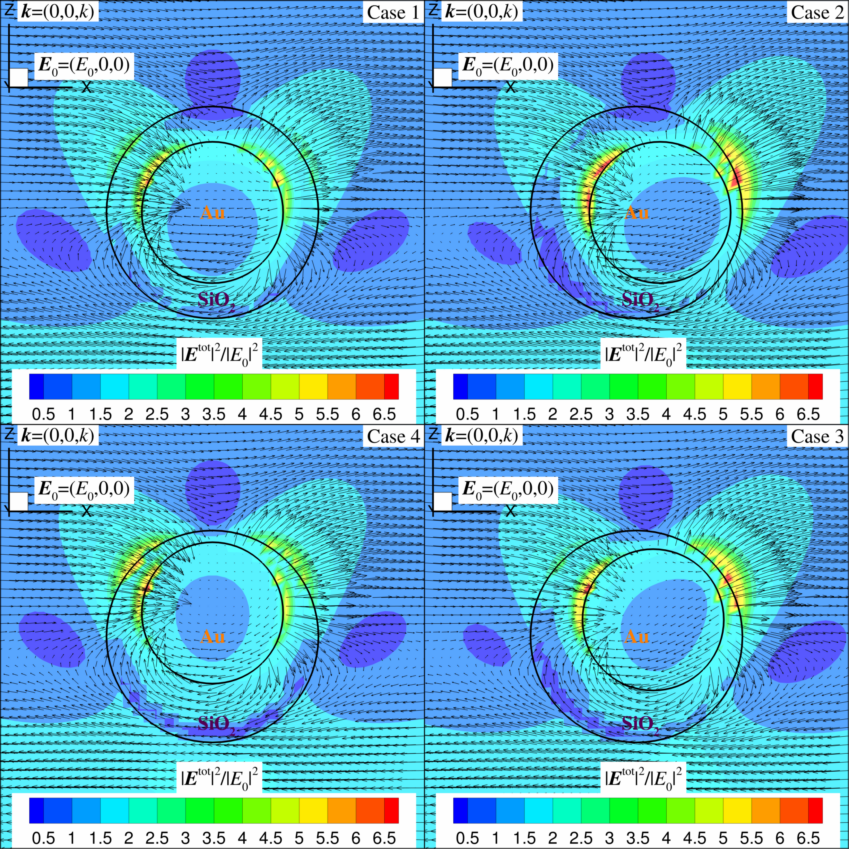} \label{Fig:CoreShellCases}}
\subfloat[(b)]{\raisebox{15ex} {\includegraphics[width=0.35\textwidth]{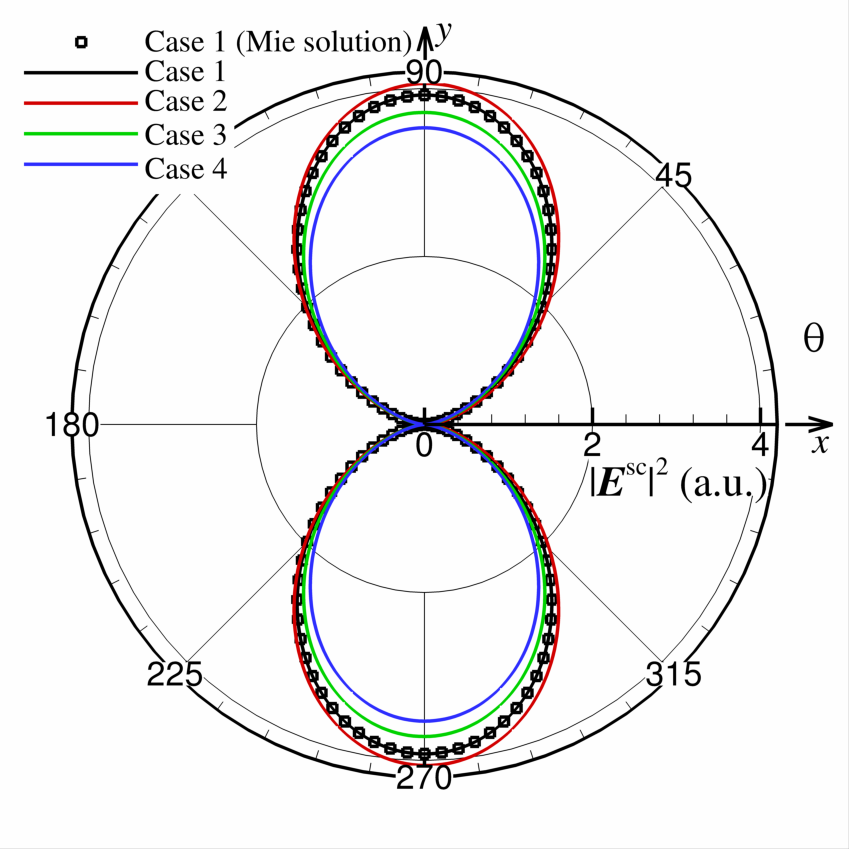}} \label{Fig:CoreShellRCS}}
\caption{(a) The total electric field vectors and contour plots of $|\bs{E}^{\mathrm{tot}}|^2/|E_0|^2$ in the $y=0$ plane are shown for four cases when the center of the core is (1) concentric with the shell, (2) shifted by $(\unit[20]{nm}) \bs{e}_x$, (3) shifted by $(\unit[14.14]{nm})\bs{e}_x + (\unit[14.14]{nm}) \bs{e}_z$, and (4) shifted by $(\unit[20]{nm})\bs{e}_z$. (b) The angular scattering intensity for the above four cases in the $z=0$ plane is shown.  For the concentric case, the corresponding Mie series solution is also shown for comparison. (Cases 2 and 3 are shown in \textcolor{urlblue}{Visualization 1}.) \label{Fig:CoreShell}} 
\end{figure*}
We illustrate the developed framework with several carefully chosen examples that show the interaction between different types of scatterers and the incident wave.  These examples are: 
\begin{enumerate}
\item Scattering by an $\mathrm{Au}$ nano-sphere located at different positions inside a $\mathrm{SiO}_{2}$ shell whose size is comparable to the wavelength of the incident wave, i.e., when the wavenumber $k$ times the characteristic size $a$ of the scatterer is of order one, $ka \sim O(1)$. The numerical procedure for this core-shell particle case is slightly more complicated due to the presence of multiple domains. When $\bs{r}_0$ is located on $S_{\text{SiO}_{2}}$, the $\Sigma$ in (\ref{eq:NS_BIE_sigma}) for the exterior domain is $\Sigma = S_{\text{SiO}_{2}} + S_{\infty}$ and $\Sigma = S_{\text{SiO}_{2}} + S_{\text{Au}}$ for the interior domain. When $\bs{r}_0$ is located on $S_{\text{Au}}$, the $\Sigma$ in (\ref{eq:NS_BIE_sigma}) for the exterior domain is $\Sigma = S_{\text{SiO}_{2}} + S_{\text{Au}}$ and $\Sigma = S_{\text{Au}}$ for the interior domain. This example illustrates the Mie scattering regime (optical wave phenomena) and is of interest, for example, in light absorption  enhancement applications for thin film solar cells~\cite{Yu2017};
\item Scattering of visible light by two $\mathrm{Au}$ nano-particles with different shapes but having the same volume.  This example shows how the shape can be used to tune the resonance wavelength and the absorption cross-section when the characteristic length of the particle is much smaller than the wavelength of the incident wave.  This scattering example is in the Rayleigh scattering regime and such quasi-electrostatic scattering problems are often encountered in micro- and nano- photonics;
\item Scattering by a dielectric oblate spheroid where the scatterer acts as a lens.   In this example, the dimension of the spheroid (lens) is larger than the incident wavelength, and thus this example is approaching the geometrical optics regime. 
\end{enumerate}%
In all these examples, the incident wave is plane wave given by $\bs{E}^{\rminc} = E_0 \exp{(\rmi k z)} \bs{e}_x$.
\begin{figure*}[!t] 
\centering
\subfloat[(a)]{\includegraphics[width=0.45\textwidth]{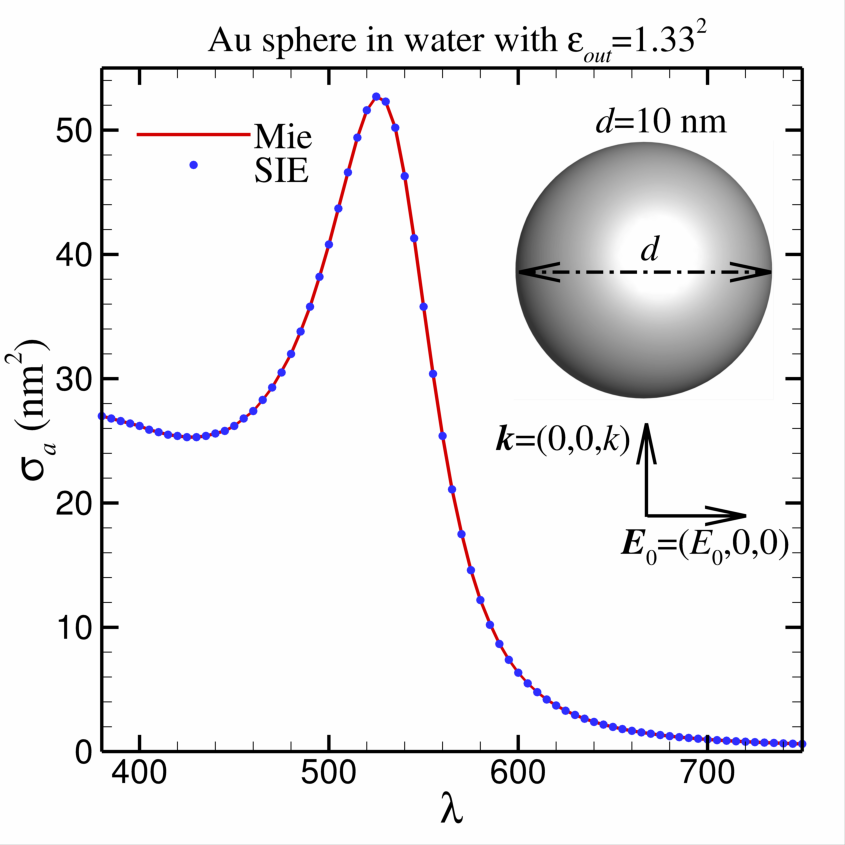} \label{fig:XSa}}
\subfloat[(b)]{\includegraphics[width=0.45\textwidth]{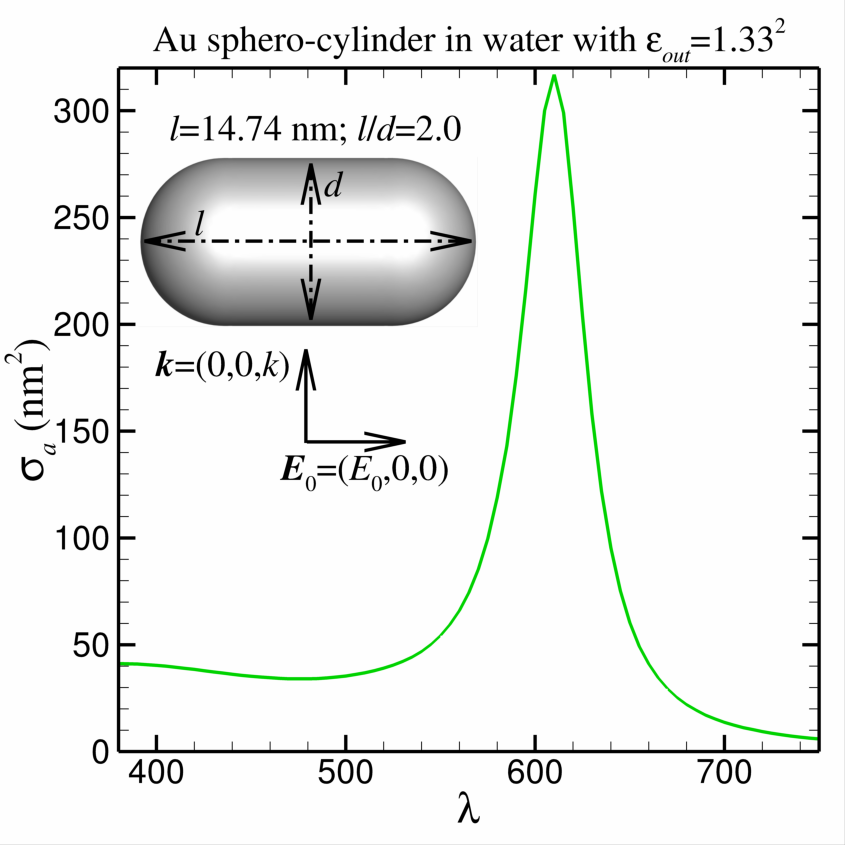} \label{fig:XSb}}
\caption{Absorption cross-section of (a) $\mathrm{Au}$ sphere and (b) $\mathrm{Au}$ sphero-cylinder (cylinder with rounded spherical sides) is shown. In (a), the continuous line is obtained from the Mie series solution while the symbols are calculated via our field-only nonsingular surface integral method. In (b), the volume of the sphero-cylinder particle is the same as the volume of the sphere in (a). \label{fig:XS}}
\end{figure*}

\subsection{Mie Scattering Example}
The scattering of a plane wave by a particle in air consisting of a metal $\mathrm{Au}$ core of radius $\unit[60]{nm}$ embedded into a $\mathrm{SiO}_2$ shell of radius $\unit[90]{nm}$ is selected as an example for Mie scattering. We chose the size of the shell to be consistent with what is used in thin film solar cells to enhance light absorption~\cite{Yu2017}.  Note that the core of the particle is not necessarily situated at the center of the shell.  The total electric field vectors and the intensity contour plots are shown in \figurename~\ref{Fig:CoreShellCases} for the incident wavelength of $\unit[520]{nm}$ (green light).   In this example, the index of refraction of the $\mathrm{SiO}_{2}$ shell is $n_{\mathrm{SiO}_{2}}=1.47$~\cite{Marcos2016} and the index of refraction of the $\mathrm{Au}$ core is $n_{\mathrm{Au}}=0.65 + 2.02 \rmi$~\cite{Rakic1998}. The plots in \figurename~\ref{Fig:CoreShellCases} are shown in the $y=0$ plane with the center of the core (1) concentric with the shell, (2) shifted by $(\unit[20]{nm}) \bs{e}_x$, (3) shifted by $(\unit[14.14]{nm})\bs{e}_x + (\unit[14.14]{nm}) \bs{e}_z$, and (4) shifted by $(\unit[20]{nm})\bs{e}_z$. 

From the electric field vector plots in \figurename~\ref{Fig:CoreShellCases}, we see that the electric fields in the $\mathrm{Au}$ core are obviously out of phase to those in the $\mathrm{SiO}_{2}$ shell. The angular scattering intensities in the $z=0$ plane are shown for all four cases in \figurename~\ref{Fig:CoreShellRCS}.  From \figurename~\ref{Fig:CoreShellRCS}, we also see that our numerical results agree very well with the Mie series solution~\cite{Mie1908, Bohren1983}.  The maximum relative difference between the two solutions is less than than $0.6\%$. 
\begin{figure*}[!t] 
\centering
\subfloat[(a)]{\includegraphics[width=0.45\textwidth]{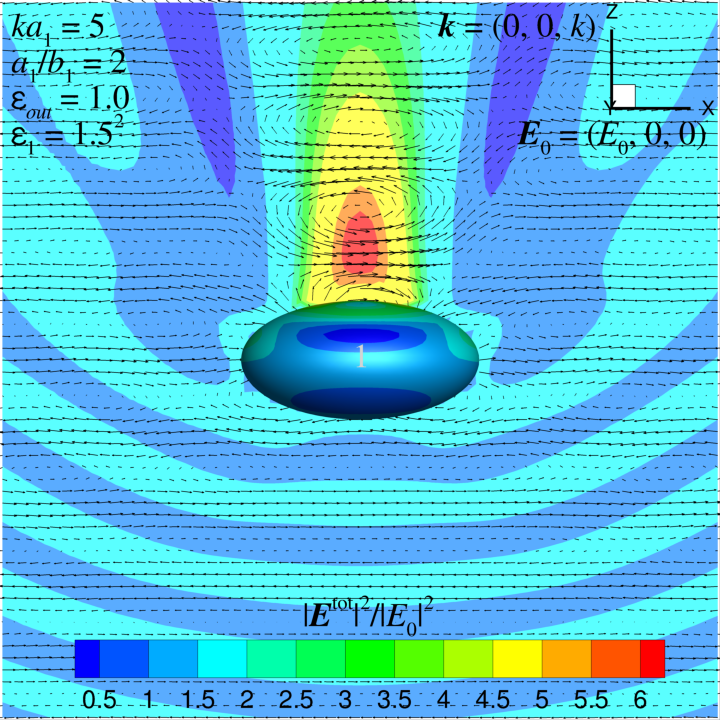} \label{fig:lens1a}}
\subfloat[(b)]{\includegraphics[width=0.45\textwidth]{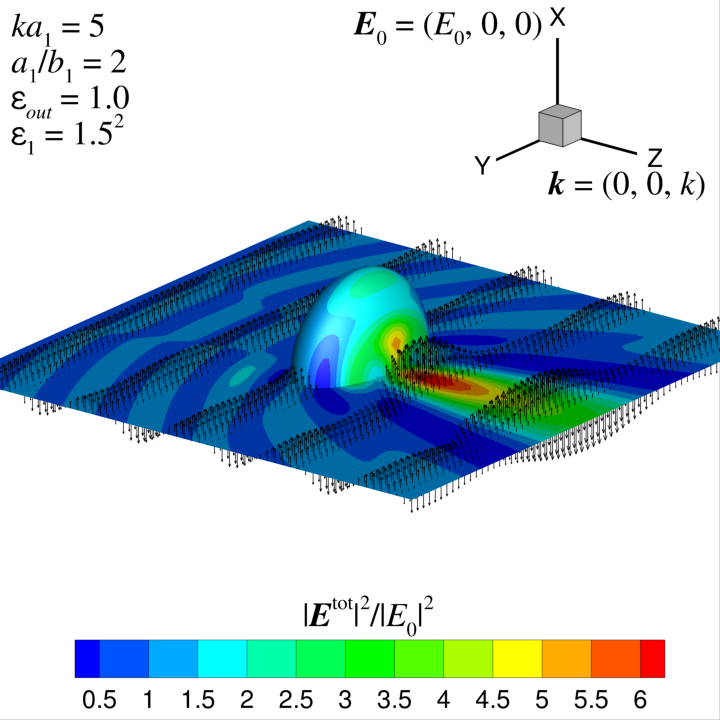} \label{fig:lens1b}}
\caption{Contour plots of $|\bs{E}^\rmtot|^2/|E_0|^2$ (a) in the $y=0$ plane, (b) in the $z=0$ plane, for a lens-shaped object approaching the geometrical optics regime. The instantaneous electric field vectors are also indicated. See also \textcolor{urlblue}{Visualization 2}.}\label{fig:lens1}
\end{figure*}

\subsection{Rayleigh Scattering Example}
In the previous example, we illustrated the ability and the accuracy of our field-only nonsingular surface integral method to solve scattering problems in the Mie scattering regime, $ka \sim O(1)$. When $ka$ is close to zero, that is, within the quasi-electrostatic limit, the scattering problem enters the Rayleigh scattering regime. The Rayleigh scattering regime is widely observed in microphotonics and nanophotonics with broad applications such as sensing of chemical and biological species~\cite{Mayer2011}. We present the effect of the scatterer's shape at a fixed volume on the absorption cross-section $\sigma_{a}$ of an $\mathrm{Au}$ particle in water with the refractive index of $n_{\mathrm{H_20}} = 1.33$. The $\mathrm{Au}$ particle is illuminated by the plane wave of wavelength varying from $\unit[380]{nm}$ to $\unit[750]{nm}$ in steps of $\unit[5]{nm}$. The absorption cross-section is calculated from the time-average Poynting vector via
\begin{equation}
  \sigma_{a} =\left(\frac{1}{I^\rminc}\right)
  \frac{1}{2} \int [\bs{E}^\rmtot \times (\bs{H}^\rmtot)^{*}]
  \cdot \rmd \bs{S},
\end{equation}
where
$I^\rminc = (1/2) v_{p} \epsilon_{0} |{E}_{0}|^2$, $v_{p}$ is the speed of the electromagnetic wave in water, and $^*$ denotes the complex conjugate. From \figurename~\ref{fig:XSa}, we see that for the spherical $\mathrm{Au}$ particle of diameter $d=\unit[10]{nm}$, the resonant wavelength occurs at around \unit[540]{nm}.  The complex index of refraction on resonance is $n_{\mathrm{Au}}=0.48 + 2.23\rmi$~\cite{Rakic1998}. Once again, the results produced via our field-only nonsingular surface integral method and the Mie series~\cite{Mie1908, Bohren1983} are in excellent agreement. Note that, even though in this example the ratio $d/\lambda \ll 1$, our method is not affected by any zero-frequency numerical instability issues.

Consider next a nano sphero-cylinder $\mathrm{Au}$ particle with the same volume as the sphere above. The length of the sphero-cylinder is $l=\unit[14.74]{nm}$ and the aspect ratio between its length and width is $l/d = 2$. From \figurename~\ref{fig:XSb}, we can see that when the long axis of the sphero-cylinder is orientated along the polarization direction of the incident wave, the resonance wavelength is red shifted to $\unit[610]{nm}$.  The complex index of refraction on resonance is $n_{\mathrm{Au}}=0.22 + 3.02\rmi$~\cite{Rakic1998} and the peak absorption cross-section is enhanced almost 6-fold relative to the $\mathrm{Au}$ sphere with the same volume. 

\subsection{Nano Lens Example}
In the previous two examples, we tested our method in the Mie and Rayleigh scattering regimes.  We now turn our attention to the geometrical optics regime, where $ka > O(1)$. Consider a dielectric oblate spheroid $(x^2+y^2)/a^2+z^2/b^2=1$ with aspect ratio of $a/b=2$ and length of $ka=5$.  The spheroid is characterized by the index of refraction of $n_{\rmin}=1.5$ and is suspended in air with the short axis parallel to the polarization of the incident wave.  From \figurename~\ref{fig:lens1}, we see that the wave is focused after it passes through the oblate spheroid and thus indicating the focusing ability of the oblate spheroid that is similar to an optical lens.  The values of the fields in \figurename~\ref{fig:lens1} were obtained via our method by first solving \eqref{eq:big_linear_system_full} and then using the method described by Sun \textit{et al.}\cite{Sun2015} to compute the field inside and outside of the oblate spheroid.  Note that the accuracy of this method is not affected by the near singular nature of the kernels when the observation point is near the boundary. If we were to use a conventional approach with its near singular Green's function based kernels, then obtaining these values would have been numerically challenging.

\section{\label{sec:CONCLUDE} Conclusions}
The electric field on, near, inside and far away from a dielectric scatterer can be obtained easily with the proposed surface integral method.  The solution satisfies both the vector Helmholtz equation and the divergence-free constraint inside and outside the scatterer. The accuracy of the solution is improved by employing a fully desingularized surface integral method. 

Some typical numerical examples were chosen representative of nano and micro optical systems. A dielectric scattering sphere was extensively tested and compared against classical Mie theory and a (nano) lens was also considered.

In our previous publication on scattering from PEC bodies~\cite{SunFuturePartI}, we listed a number of advantages of our surface integral method over the most popular methods based on the Stratton--Chu\cite{Stratton1941, Stratton1939} or the PMCHWT\cite{Poggio1973, Wu1977, Chang1977} formulation.  In this paper, we have shown that these advantages carry over to the dielectric case.  For completeness and ease of reference we list these advantages here once more; namely,
\begin{enumerate}
    \item Our method is conceptually simple and numerically straightforward because it focuses on solving directly for physically important quantities, namely, the electric field and its normal derivative on the surface of the scatterer. One of most obvious application of this method is its usefulness in computing the optical force on the dialectic particles;
    \item Our method does not need to work with intermediate quantities such as surface currents. As such, elaborate vector basis functions (such as RWG~\cite{BelezJOSA2015, LiJOSA2014}) are not required and the standard boundary element techniques can be employed. Furthermore, our method only requires a boundary element solver for the scalar Helmholtz equation;
    \item The robust, effective and accurate nonsingular surface integral method~\cite{Klaseboer2012, Sun2015} (also see \ref{Appendix_desingularized}) that is based on nonsingular integrands and uses quadratic surface elements provides a more precise representation of the boundary geometry;
    \item Our method may be advantageous in solving time-domain scattering problems using inverse Fourier transforms~\cite{Klaseboer2017_2} because it directly solves for the electric field;
    \item The framework presented here is not affected by certain numerical issues encountered in other implementations.  For example, there are no integrals with strong singularities~\cite{Chew1989} and the zero frequency catastrophe~\cite{Vico2016, Zhao2000} is avoided altogether. 
\end{enumerate}

Given the symmetry between the $\bs{E}$ field and the $\bs{H}$ field, our theoretical framework can also be used to solve for the $\bs{H}$ field. One only needs to replace $\bs{E}$ by $\bs{H}$, $\bs{H}$ by $-\bs{E}$, and interchange $\epsilon$ with $\mu$ in the formulas given above.


\renewcommand{\thesection}{Appendix A}
\section{Continuity of H-Field on the Interface\label{Appendix_HnHt}}
\setcounter{equation}{0}
\renewcommand{\theequation}{A\arabic{equation}}%
The tangential component of $\bs{H}$ in the $\bs{t}_1$-direction can be expressed as $H_{t_1}=\bs{t}_1 \cdot (\bs{n} \times \bs{H})$. The magnetic field can be expressed in terms of the electric field via $(\omega \mu) \bs{H} = -\rmi \bs{\nabla} \times \bs{E} $. Thus, for the tangential component of the magnetic field we have
\begin{align} 
\left(\frac{\omega \mu}{\rmi}\right)H_{t_1}=-\bs{t}_1 \cdot (\bs{n} \times \bs{\nabla} \times \bs{E})  
=  \bs{t}_1 \cdot \frac{\partial \bs{E}}{\partial n} - \bs{n} \cdot \frac{\partial \bs{E}}{\partial t_1}.\label{eq:Ht3}
\end{align}
Writing $\partial \bs{E}/\partial t_1$ as 
\begin{align}\label{eq:dEdt1}
 \frac{\partial \bs{E}}{\partial t_1} \equiv& 
 \frac{\partial  (E_n \bs{n} + E_{t_1} \bs{t}_1 + E_{t_2} \bs{t}_2)}{\partial t_1} \nonumber \\
 =&\quad \bs{n} \frac{\partial E_n}{\partial t_1} +E_n  \frac{\partial \bs{n}}{\partial t_1} + \bs{t}_1 \frac{\partial E_{t_1}}{\partial t_1}+E_{t_1}  \frac{\partial \bs{t}_1}{\partial t_1} \nonumber 
 +\bs{t}_2 \frac{\partial E_{t_2}}{\partial t_1} +E_{t_2}  \frac{\partial \bs{t}_2}{\partial t_1}
\end{align}
using $\partial \bs{n}/\partial t_1 = -\kappa_1 \bs t_1$ and $\partial \bs{t}_1/\partial t_1 = \kappa_1 \bs n$, with $\kappa_1$ the curvature in the $t_1$ direction (see identities (A9)--(A11) in \cite{SunFuturePartI}) yields
\begin{equation} \label{eq:dEdt2}
 \bs{n} \cdot \frac{\partial \bs{E}}{\partial t_1} = \frac{\partial E_n}{\partial t_1} + \kappa_1 E_{t_1}.
\end{equation}
Applying \eqref{eq:Ht_BC} and \eqref{eq:Ht3} to the incident, scattered, and transmitted fields yields
\begin{equation} \label{eq:Ht2}
\begin{aligned}
&\frac{1}{\mu_{\rmout}}\left[\bs{t}_1 \cdot \frac{\partial (\bs{E}^\rminc + \bs{E}^\rmsc)}{\partial n} - \bs{n} \cdot \frac{\partial (\bs{E}^\rminc + \bs{E}^\rmsc)}{\partial t_1}\right]\\
=&\frac{1}{\mu_{\rmin}}\left[\bs{t}_1 \cdot \frac{\partial \bs{E}^\rmtr}{\partial n} - \bs{n} \cdot \frac{\partial \bs{E}^\rmtr}{\partial t_1}\right]
\end{aligned}
\end{equation}
and, after using \eqref{eq:dEdt2}, we obtain the desired result \eqref{eq:dEdn_t_BC}. A similar derivation can be done for the tangential component in the $\bs{t}_2$-direction.

\renewcommand{\thesection}{Appendix B}
\section{Nonsingular Surface Integral Equation \label{Appendix_desingularized}}
\setcounter{equation}{0}
\renewcommand{\theequation}{B\arabic{equation}}%
A brief description of the nonsingular surface integral method to solve the scalar Helmholtz equation is now presented. Take a scalar function $p(\bs{r})$ that satisfies the 3D Helmholtz equation $\nabla^2 p(\bs{r}) + k^2 p(\bs{r}) = 0$, where, for example, $p$ represents one of the Cartesian components of the electric field. The nonsingular surface integral equation is given by~\cite{Klaseboer2012}%
\begin{align} \label{eq:NS_BIE}
&\int_{S} {\Big[p(\bs{r}) - p(\bs{r}_0) g(\bs{r}) - \frac{\partial {p(\bs{r}_0)}} {{\partial {n}}}  f(\bs{r})\Big] \frac{\partial {G}} {{\partial {n}}} \, \rmd S(\bs{r}}) = \nonumber \\  &\int_{S} { \Big[\frac{\partial {p (\bs{r})}} {{\partial {n}}} - p(\bs{r}_0) \frac{\partial {g (\bs{r})}} {{\partial {n}}} - \frac{\partial {p(\bs{r}_0)}} {{\partial {n}}} \frac{\partial {f (\bs{r})}} {{\partial {n}}} \Big] G \, \rmd S(\bs{r}}), 
\end{align} %
where $\bs{r}$ is the source point and $\bs{r}_0$ is the field (observation) point. The functions $f(\bs{r})$ and $g(\bs{r})$ in  \eqref{eq:NS_BIE} must satisfy the Helmholtz equation and also satisfy the following conditions at $\bs{r} =\bs{r}_0$:
\begin{subequations} \label{eq:fg_constraints}
\begin{eqnarray}
f(\bs{r}_0) &=& 0,  \qquad   \bs{n}(\bs{r}_0) \cdot  \bs{\nabla} f(\bs{r}_0) = 1,
\\
g(\bs{r}_0) &=& 1, \qquad \bs{n}(\bs{r}_0) \cdot \bs{\nabla} g(\bs{r}_0) = 0.
\end{eqnarray}
\end{subequations}
The functions $f(\bs{r})$ and $g(\bs{r})$ are not uniquely determined,  see Klaseboer {\emph{et al.}}~\cite{Klaseboer2012} and Sun {\emph{et al.}}~\cite{Sun2015} for more details. Note that the solid angle will be eliminated using this framework. In this paper, we used two standing wave functions for $f$ and $g$, i.e.,  
\begin{subequations}\label{eq:f_g}
\begin{eqnarray} 
f(\bs{r}) &=& \displaystyle{\frac{1}{k}} \sin \big( k \bs{n}(\bs{r}_0) \cdot [\bs{r} - \bs{r}_0] \big),\\
g(\bs{r}) &=& \cos \big( k \bs{n}(\bs{r}_0) \cdot [\bs{r} - \bs{r}_0] \big).
\end{eqnarray}
\end{subequations}
If the Helmholtz equation is solved in the domain exterior to the scatterer, an additional factor $4\pi p(\bs{r}_0)$ must be added to the left hand side of \eqref{eq:NS_BIE} due to the particular choice we made in \eqref{eq:f_g}.  This contribution results from evaluating \eqref{eq:NS_BIE} over a fictitious surface at infinity and depends on the choice of $f$ and $g$.  Equation~\eqref{eq:NS_BIE} is essentially the standard boundary element method implementation, where a known analytic solution $p(\bs{r}_0) g(\bs{r}) + [\partial p(\bs{r}_0)/\partial n] f(\bs{r})$ has been subtracted. In this context, $p(\bs{r}_0)$ and $\partial p(\bs{r}_0)/\partial n$ are constants (for one particular node $\bs{r}_0$). This framework has been extensively tested for the Helmholtz equation in sound waves~\cite{Sun2015}, electromagnetic scattering~\cite{Sun2017}, and even elastic waves in solids~\cite{Klaseboer2018}. Due to the fact that the formulation is nonsingular, Gaussian quadrature can be used on all elements (including the previous singular ones) and the implementation of higher order elements is straightforward. 

Although the desingularized surface integral framework is not essential to solve the considered electromagnetic scattering problem, it does greatly improve the ease of implementation of the matrix systems $\cal{G}$ and $\cal{H}$, see \eqref{eq:BIE_GH_Matrix} and Appendix~B in~\cite{SunFuturePartI}, and it also improves the accuracy of the solution.

\renewcommand{\thesection}{Appendix C}
\section{Partial Derivatives Matrix \label{Appendix_dtMatrix}} %
\setcounter{equation}{0}
\renewcommand{\theequation}{C\arabic{equation}}%
\begin{figure}
  \centering
    \includegraphics[width=0.35\textwidth]{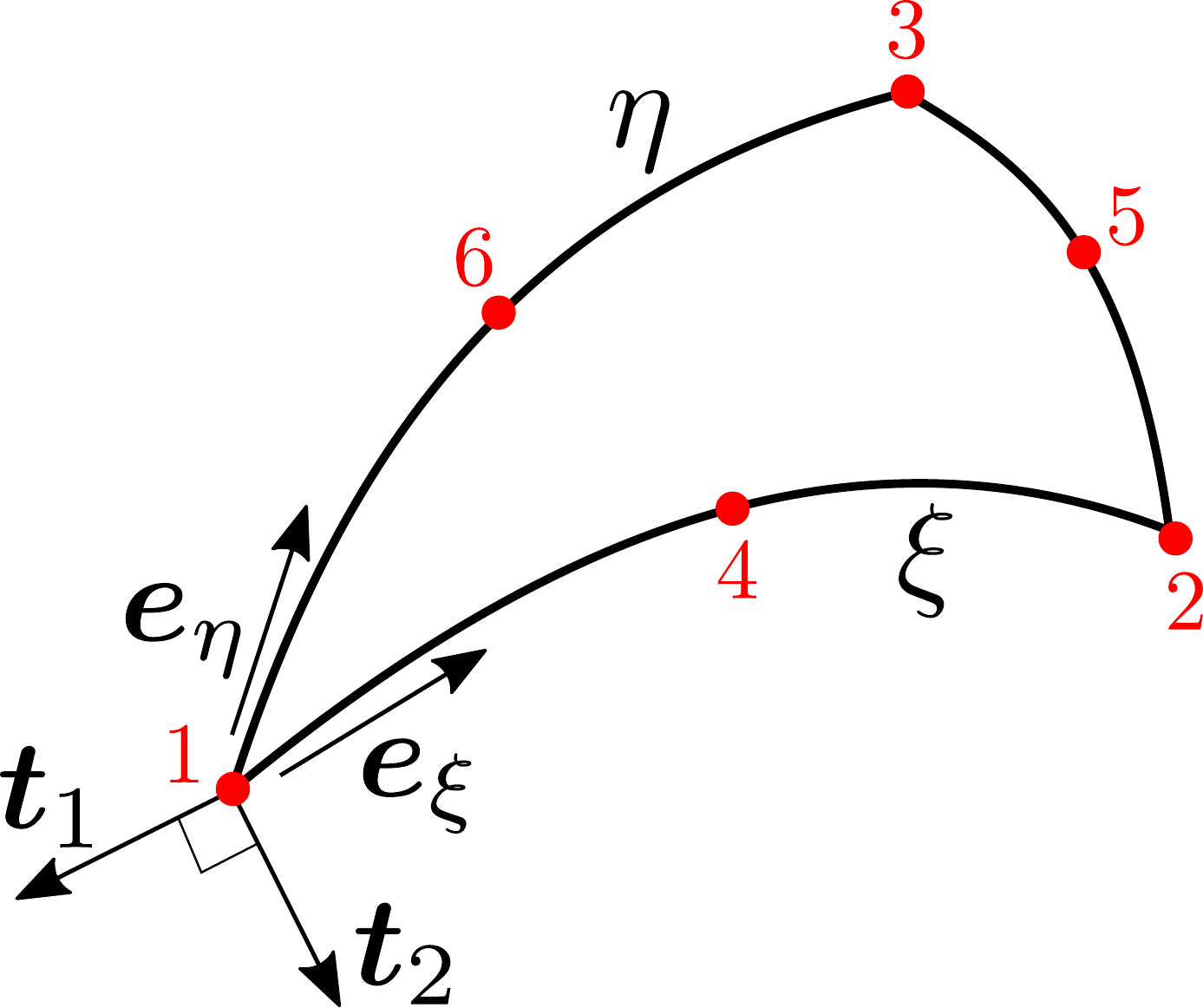}
    \caption{A quadratic surface patch with six nodes is shown.\label{Fig:dt1dt2}}
\end{figure}
The tangential derivatives, $\partial /\partial t_1$ and $\partial /\partial t_2$, can be obtained from the directional derivatives. Suppose that the tangential derivatives are sought at node $1$ of the six-noded quadratic surface element shown in \figurename~\ref{Fig:dt1dt2}. Assume that the side 1-4-2 is represented by $\xi$ and the side 1-6-3 by $\eta$. Unit vectors in these two directions are denoted by $\bs{e}_\xi$ and $\bs{e}_\eta$, respectively. While $\bs{t}_1$ and $\bs{t}_2$ are perpendicular to each other, in general, $\bs{e}_\xi$ and $\bs{e}_\eta$ are not. The directional derivative in the $\xi$-direction can be written as
\begin{align} 
D_{\xi} f &=\bs{e}_\xi \cdot \bs{\nabla} f = \bs{e}_\xi \cdot \left[\bs{t}_1 \frac{\partial f}{\partial t_1} + \bs{t}_2 \frac{\partial f}{\partial t_2} + \bs{n} \frac{\partial f}{\partial n}\right] \nonumber \\
&\approx \frac{f_4-f_1}{l_{41}}, \label{eq:dfdxi}
\end{align}
where $l_{41}$ denotes the Euclidean distance between nodes $1$ and $4$.  Note that to obtain \eqref{eq:dfdxi}, we used a simple numerical approximation to the directional derivative. Similarly, for the $\eta$-direction we use nodes $6$ and $1$ to obtain
\begin{equation}\label{eq:dfdeta}
D_{\eta} f = (\bs{e}_\eta \cdot \bs{t}_1) \frac{\partial f}{\partial t_1} + (\bs{e}_\eta \cdot \bs{t}_2) \frac{\partial f}{\partial t_2} \approx \frac{f_6-f_1}{l_{61}}.
\end{equation}
Solving \eqref{eq:dfdxi} and \eqref{eq:dfdeta} yields
\begin{subequations}\label{eq:partials_from_nodes}
\begin{align}
\frac{\partial f}{\partial t_1} &=\frac{1}{D}\left[(\bs{e}_\xi \cdot \bs{t}_2) \frac{f_6 - f_1}{l_{61}} - (\bs{e}_\eta \cdot \bs{t}_2) \frac{f_4-f_1}{l_{41}} \right] \\
\intertext{and}
\frac{\partial f}{\partial t_2} &=\frac{1}{D}\left[(\bs{e}_\eta \cdot \bs{t}_1) \frac{f_4 - f_1}{l_{41}} - (\bs{e}_\xi \cdot \bs{t}_1) \frac{f_6-f_1}{l_{61}} \right], 
\end{align}
\end{subequations}
where the determinant $D=(\bs{e}_\xi \cdot \bs{t}_2) (\bs{e}_\eta \cdot \bs{t}_1) - (\bs{e}_\xi \cdot \bs{t}_1) (\bs{e}_\eta \cdot \bs{t}_2)$.  Notice that \eqref{eq:partials_from_nodes} expresses $\partial f/ \partial t_1$ and $\partial f/ \partial t_2$ in terms of the values at the nodes . The simplest possible implementation is shown above, but a more accurate quadratic scheme can be obtained by using nodes 1, 4 and 2 in the numerical derivatives for $\xi$. As a further improvement, the above scheme has been applied to all elements surrounding node 1 and was averaged by the number of surrounding elements. The implementation for other nodes is very similar. In the numerical implementation, $f$ is the unknown variable $E_n^{\rmsc}$ and thus, $\partial / \partial t_1$ and $\partial /\partial t_2$ become $N \times N$ matrices, which have to be multiplied with the matrix $t_{1\alpha} {\cal{G}}_{\rmin}$ to contribute to the terms $\bar{{\cal{H}}}^{n\alpha}_{\rmin}$ in \eqref{eq:big_linear_system_full}. 

The matrices representing $\partial / \partial t_1$ and $\partial /\partial t_2$ can also elegantly be employed to calculate the curvatures $\kappa_1$ and $\kappa_2$ via (see (A9c) and (A9d) in \cite{SunFuturePartI})
\begin{equation}
\kappa_1  = - \bs{t}_{1} \cdot \frac{\partial \bs{n}}{\partial t_1} \quad \text{and} \quad \kappa_2  = - \bs{t}_{2} \cdot \frac{\partial \bs{n}}{\partial t_2}.
\end{equation}

\renewcommand{\thesection}{Appendix D}
\section{Linear Matrix System\label{Appendix_MatrixEtr}} %
\setcounter{equation}{0}
\renewcommand{\theequation}{D\arabic{equation}}%
We now demonstrate how to assemble the linear matrix system of equations in terms of the interior field $\bs{E}^{\rmtr}$ and $\partial \bs{E}^{\rmtr}/\partial n $ following the same procedure given in Section 2.

To write $\bs{n} \cdot \partial \bs{E}^\rmsc / \partial n$ in terms of $\bs{E}^{\rmtr}$ and $\partial \bs{E}^{\rmtr}/\partial n $, we can rearrange (\ref{eq:dEdn_n_BC}), and after using \eqref{eq:En_BC}, we have
\begin{equation} \label{eq:dEdn_n_BC2}
  \bs{n} \cdot \frac{\partial { \bs{E}^{\rmsc} }  }{ \partial{n} } =  \kappa (\epsilon_{\rmio}-1) E^{\rmtr}_{n} +  \bs{n} \cdot \frac{\partial { \bs{E}^{\rmtr} }  }{ \partial{n} } - \bs{n} \cdot \frac{\partial { \bs{E}^{\rminc} }  }{ \partial{n} } \qquad \text{where} \qquad \epsilon_{\rmio} \equiv \epsilon_{\rmin}/\epsilon_{\rmout}.
\end{equation}

To write $\{\bs{t}_1 \cdot \partial \bs{E}^\rmsc/ \partial n,\, \bs{t}_2 \cdot \partial \bs{E}^\rmsc/ \partial n\}$ in terms of $\bs{E}^{\rmtr}$ and $\partial \bs{E}^{\rmtr}/\partial n $ we express the continuity of the tangential components of $\bs{H}$ on $S$ given by \eqref{eq:Ht_BC} in terms of the electric field with $\mu_{\rmoi} \equiv \mu_{\rmout}/\mu_{\rmin}$ and $j=1,2$ (see \ref{Appendix_HnHt} for details):
\begin{multline}\label{eq:dEdn_t_BC2}
  \left(\epsilon_{\rmio}- \mu_{\rmoi} \right) \frac{\partial}{\partial t_j} E_n^{\rmtr}
  + \kappa_j\left(1-\mu_{\rmoi}\right)  E_{t_j}^{\rmtr} - \bs{t}_j \cdot \frac{\partial \bs{E}^{\rminc}}{\partial n} 
  +\mu_{\rmoi} \bs{t}_j \cdot \frac{\partial \bs{E}^{\rmtr} }{\partial n} 
  =  \bs{t}_j \cdot \frac{\partial \bs{E}^{\rmsc}}{\partial n}.
\end{multline}
As such, the linear system in terms of $\bs{E}^{\rmtr}$ and $\partial \bs{E}^{\rmtr}/\partial n $ can be found to be
%
\begin{subequations}\label{eq:big_linear_system_fullEtr}
  \begin{equation}\label{eq:big_linear_systemEtr}
    \begin{bmatrix}
      \bar{{\cal{H}}}^{nx} & \bar{{\cal{H}}}^{t1x} & \bar{{\cal{H}}}^{t2x}
      & - {n_{x}}{\cal{G}} & -\mu_{\rmoi} {t_{1x}}{\cal{G}}
      & -\mu_{\rmoi} {t_{2x}}{\cal{G}} \\
      \bar{{\cal{H}}}^{ny} & \bar{{\cal{H}}}^{t1y}
      & \bar{{\cal{H}}}^{t2y}  & - {n_{y}}{\cal{G}}
      & -\mu_{\rmoi} {t_{1y}}{\cal{G}}  & -\mu_{\rmoi} {t_{2y}}{\cal{G}} \\
      \bar{{\cal{H}}}^{nz} & \bar{{\cal{H}}}^{t1z} & \bar{{\cal{H}}}^{t2z}
      & - {n_{z}}{\cal{G}} & -\mu_{\rmoi} {t_{1z}}{\cal{G}}
      & -\mu_{\rmoi} {t_{2z}}{\cal{G}} \\
       n_{x} {\cal{H}}_{\rmin}  & t_{1x} {\cal{H}}_{\rmin}
      &   t_{2x} {\cal{H}}_{\rmin}  & -n_{x} {\cal{G}}_{\rmin}
      & - t_{1x} {\cal{G}}_{\rmin} & -t_{2x} {\cal{G}}_{\rmin}  \\
       n_{y} {\cal{H}}_{\rmin}  & t_{1y} {\cal{H}}_{\rmin}
      &  t_{2y}  {\cal{H}}_{\rmin}  & -n_{y} {\cal{G}}_{\rmin}
      & -t_{1y} {\cal{G}}_{\rmin}  & -t_{2y} {\cal{G}}_{\rmin}  \\ 
       n_{z} {\cal{H}}_{\rmin}  & t_{1z} {\cal{H}}_{\rmin}
      &  t_{2z} {\cal{H}}_{\rmin}   & -n_{z} {\cal{G}}_{\rmin}
      & -t_{1z} {\cal{G}}_{\rmin}  & -t_{2z}  {\cal{G}}_{\rmin} 
    \end{bmatrix}
    \begin{bmatrix}
      E_n^{\rmtr} \\ E_{t_{1}}^{\rmtr} \\ E_{t_{2}}^{\rmtr} \\
      \bs{n} \cdot \frac{\partial \bs{E}^{\rmtr}}{\partial n}   \\
      \bs{t}_{1} \cdot \frac{\partial \bs{E}^{\rmtr}}{\partial n}  \\
      \bs{t}_{2} \cdot \frac{\partial \bs{E}^{\rmtr}}{\partial n}
    \end{bmatrix}
    =
    \begin{bmatrix}
      {\cal{A}}_{x}\\ {\cal{A}}_{y} \\ {\cal{A}}_{z} \\
      0 \\ 0 \\ 0
    \end{bmatrix}, 
  \end{equation}
  where
  \begin{align}
    \bar{{\cal{H}}}^{n\alpha} &= \epsilon_{\rmio}n_{\alpha} {\cal{H}}
                                - \kappa \left(\epsilon_{\rmio}-1\right)
                                n_{\alpha} {\cal{G}}
                                - (\epsilon_{\rmio}-\mu_{\rmoi})
                                t_{1\alpha}{\cal{G}}
                                \frac{\partial }{\partial t_1}
                                - (\epsilon_{\rmio}-\mu_{\rmoi})
                                t_{2\alpha}  {\cal{G}}
                                \frac{\partial}{\partial t_2}, \label{eq:matrix_patial_derivativeEtr} \\
    \bar{{\cal{H}}}^{t1\alpha} &= t_{1\alpha} {\cal{H}}
                                 -  \kappa_{1}(1-\mu_{\rmoi})
                                 t_{1\alpha}  {\cal{G}} \quad \text{and} \quad
                                 \bar{{\cal{H}}}^{t2\alpha} =   t_{2\alpha}
                                 {\cal{H}} - \kappa_{2} (1-\mu_{\rmoi})
                                 t_{2\alpha}  {\cal{G}},
  \end{align}
  and
  \begin{align}
    {\cal{A}}_{\alpha} =&\;\;\;  {\cal{H}} \left[ n_{\alpha} E_n^{\rminc} + t_{1\alpha} E^{\rminc}_{t_{1}} + t_{2\alpha}E^{\rminc}_{t_{2}}\right]  \nonumber \\
                        &- {\cal{G}} \left[n_{\alpha}
                          \bs{n} \cdot
                         \frac{\partial \bs{E}^{\rminc}}{\partial n}
                         + t_{1\alpha}
                         \bs{t}_1 \cdot
                         \frac{\partial \bs{E}^{\rminc}}{\partial n} 
                          + t_{2\alpha}
                           \bs{t}_{2} \cdot
                          \frac{\partial \bs{E}^{\rminc}}{\partial n}\right]
  \end{align} 
  with $\alpha = x,y,z.$
\end{subequations}
%


\section*{Funding}
Australian Research Council (ARC) (DE150100169,  CE140100003, DP170100376).\\
This work was partially supported by U.S.~government, not protected by U.S.~copyright.


\section*{Disclosures}
The authors declare that there are no conflicts of interest related to this article.

 
\bibliography{refs_dielectric}

\end{document}